\documentclass[acmsmall,authorversion]{acmart}

\renewcommand\footnotetextcopyrightpermission[1]{} %

\usepackage[toc,page]{appendix}
\usepackage{hyperref}
\usepackage[table]{xcolor}
\usepackage{pifont}
\usepackage{graphicx}
\usepackage{subcaption}
\usepackage[nolist]{acronym}
\usepackage{cleveref} %
\usepackage{soul} %
\usepackage{balance}

\usepackage{adjustbox}
\usepackage{array}

\newcolumntype{R}[2]{%
    >{\adjustbox{angle=#1,lap=\width-(#2)}\bgroup}%
    l%
    <{\egroup}%
}
\newcommand*\rot{\multicolumn{1}{R{45}{1em}}}%

\makeatletter
\AtBeginDocument
 {
   \def\ltx@label#1{\cref@label{#1}}%
   \def\label@in@display@noarg#1{\cref@old@label@in@display{#1}}%
\def\label@in@mmeasure@noarg#1{%
    \begingroup%
      \measuring@false%
      \cref@old@label@in@display{#1}%
    \endgroup}%
 } %
\makeatother

\author{Moritz M\"uller-Brus}
  \affiliation{%
    \institution{SIDN Labs}
    \country{The Netherlands}
    \city{Arnhem}}
  \affiliation{%
    \institution{University of Twente}
    \country{The Netherlands}
    \city{Enschede}}
\email{moritz.muller@sidn.nl}
\author{Lisa Bruder}
  \affiliation{%
    \institution{SIDN Labs}
    \country{The Netherlands}
    \city{Arnhem}}
\email{lisa.bruderr@sidn.nl}
\author{Caspar Schutijser}
  \affiliation{%
    \institution{SIDN Labs}
    \country{The Netherlands}
    \city{Arnhem}}
\email{caspar.schutijser@sidn.nl}
\author{Ralph Koning}
  \affiliation{%
    \institution{SIDN Labs}
    \country{The Netherlands}
    \city{Arnhem}}
  \affiliation{%
    \institution{University of Amsterdam}
    \country{The Netherlands}
    \city{Amsterdam}}
\email{ralph.koning@sidn.nl}

\title{A first look at common RPKI publication practices}

\begin{abstract}
The \ac{RPKI} is crucial for securing the routing system of the Internet. With the \ac{RPKI}, owners of Internet resources can make cryptographically backed claims, for example about the legitimate origin of their IP space. Thousands of networks use this information to detect malicious or accidental route hijacks. The \ac{RPKI} consists out of 100 distributed repositories. However, public reports claim that some of these repositories are unreliable. A current Internet-Draft suggests best practices on how to operate these repositories, with the goal to improve deployment quality. 

Inspired by this draft, we take a first look at the operational practices of repositories of the \ac{RPKI}. We mainly focus on the \emph{distribution} of \ac{RPKI} information.  We find that there is a wide variety in deployment practices, of which some might risk the availability of parts of the information in the \ac{RPKI}. This study creates a baseline for measuring the maturity of \ac{RPKI} repositories in the future. 
\end{abstract}

\begin{document}

\maketitle

\section{Introduction} 

The \ac{RPKI} enables holders of Internet resources,
 like IP address space, to make cryptographically backed claims
 about their assets~\cite{rfc6480}. Currently, the \ac{RPKI} mostly contains claims
 about IP prefixes and the networks from which those prefixes may be
 announced in so called \acp{ROA}. Networks run \ac{RP} software to fetch and validate these objects and then base their \ac{BGP} routing decisions on the validation results.
At the moment, almost 60\% of all IP prefixes are covered by
 a \ac{ROA}~\cite{nist-rpki-monitor-rov-all-4},
  thereby protecting these
 prefixes against malicious attacks and misconfiguration. The validation of \acp{ROA} is also on the rise \cite{10.1145/3618257.3624806}. 

The \ac{RPKI} relies on ``a distributed repository
system for storing and disseminating'' its objects~\cite{rfc6480}. Most of these objects,
like \acp{ROA}, are managed and distributed by one of the five \acp{RIR}
(AFRINIC, APNIC, ARIN, LACNIC, RIPE NCC). At the same time, the \ac{RPKI} 
enables resources holders to operate their own \ac{CA}. Then, resource owners can decide to 
publish the objects themselves or, if provided, rely on the
publication service by their \ac{RIR}. The \ac{RIR} or the resource owner
needs to make sure that the objects can be retrieved by \acp{RP} using
the HTTP based \ac{RRDP} or rsync. 

Besides the repositories by the 5 RIRs, 95 repositories are
currently listed in the \ac{RPKI}. In recent years, operators of \ac{RP}
software frequently complained about repositories that were
unreliable~\cite{li-sidrops-rpki-repository-problem-statement-02}. 
Aside from putting extra burden on \ac{RP} 
software, unavailable repositories can also negatively impact the
security of Internet resources. If a repository is unavailable for too
long (e.g.\ several hours), objects published by it can not be updated and potentially expire,
thereby effectively removing the protection by the RPKI. In certain
scenarios, Internet resources covered by objects could even
become unavailable for (parts of) the Internet.

This, likely, motivated
 the authors of \emph{RPKI Publication Server Best Current Practices}
 to formulate a set of practices that operators of publication servers
should consider implementing. The IETF working group ``SIDR Operations'' has 
adopted this document as an Internet-Draft~\cite{ietf-sidrops-publication-server-bcp-05}.

Inspired by this effort, our goal is to measure the current \ac{RPKI} publication
practices. Thereby, we also create a baseline measurement of the
implementation of the \acp{BCP}. 
Note that the goal of this study is not find out \emph{why} some
operators run their own publication server, nor to measure the performance of
the publication servers. Our contributions are as follows:

\begin{itemize}
	\item Provide a first look at \ac{RPKI} publication practices, showing that there are only a few common practices most operators implement.
	\item Create a baseline measurement for tracking the impact of the recommended best practices.
	\item Highlight widely diverging practices between operators, showing that there is not \emph{the} best way to publish objects in the \ac{RPKI}.
\end{itemize}

\section{RPKI Basics}

In the \ac{RPKI}, signed objects are grouped together into publication points based on the signing \ac{CA}. Every group of objects is accompanied by a manifest that lists all 
signed objects and a \ac{CRL} that lists all objects that have been revoked.

The 5 \acp{RIR} serve as trust anchors in the \ac{RPKI} and the private keys associated with their \ac{CA} certificates are used to sign certificates for subordinate \acp{CA} as well as resource certificates~\cite{rpki_docs}.
Resource holders can let their \ac{RIR} manage the signing of their objects (so-called hosted \ac{RPKI}). Alternatively, it is possible to request a delegated \ac{CA} certificate 
from an \ac{RIR} and manage signing independently (delegated \ac{RPKI})~\cite{rpki_docs}.

Objects also need to be hosted for \acp{RP} to retrieve them. Objects stored in repositories in the \ac{RPKI} are distributed 
using rsync and \ac{RRDP} servers \cite{rfc8182}. When resource holders make use of the ``hosted'' RPKI service of their \ac{RIR}, the objects are hosted through the \ac{RIR}'s \ac{RRDP} and rsync servers. 
However, objects can also be distributed through self-hosted rsync and \ac{RRDP} servers \cite{rpki_docs}.

\ac{RRDP} is an HTTP based protocol. A CA certificate contains a references to a notification file, which is downloaded by an \ac{RP}, e.g.\ every 10 minutes. 
This file includes a reference to a snapshot file, which contains the full content of the repository. 
Additionally, the notification file can contain a list of delta files, listing incremental changes to the repository \cite{rfc8182}. 
An \ac{RP} checks if it can use the delta files to bring its cache up to date or if it needs to download the full snapshot file. 
If the \ac{RRDP} server is not reachable, the rsync server can be used to retrieve the objects.

If neither the \ac{RRDP} nor the rsync server are reachable for a prolonged time, no new certificates and signed objects can be distributed.
Then, the objects signed by the \ac{CA} could expire. Because current practice does not advice operators to drop routes with a status of ``unknown'', 
the impact of an expired \ac{ROA} object is limited \cite{rfc7454}. However, in specific situations where a prefix is covered by two \acp{ROA} associated with different \acp{ASN}, 
an announcement could become invalid if one of the \ac{ROA} objects expires \cite{10.1145/3603269.3604861, krill_docs}.

\section{Methodology}
\label{sec:methodology}
The Internet-Draft ``RPKI Publication Server Best Current Practices'' (I-D) \cite{ietf-sidrops-publication-server-bcp-05} describes best current
practices for the operation of a publication, rsync and \ac{RRDP} servers based on operational experience. This methodology applies to version 05 of the I-D. 
In this section, we briefly describe the goal of the recommended practices and
the measurements that we performed, grouped by the section in the I-D they
appear in. Each measurement has a short name that we use to refer to the
measurement in the paper, printed in (\textbf{bold}). We refer the reader to the I-D for a more detailed description and motivation for each recommended practice.

We use \ac{RP} software rpki-client v9.6 to download the current data published in the \ac{RPKI}. The
rpki-client cache serves as a basis for our analysis. It contains all files
that have a valid signature path. Our measurement vantage point, located in a
well connected network , supports both IPv4 and IPv6 and we
fetch the content over whatever protocol preferred by rpki-client. For
each \ac{CA} certificate we encounter in the \ac{RPKI} data we extract
the \acp{URL} of the \ac{RRDP} and rsync servers from the Subject Information
Access extension fields.\footnote{e.g.\ \url{https://rrdp.ripe.net/notification.xml} and
\url{rsync://rpki.ripe.net/repository/DEFAULT/} for the \ac{CA}
cert. of the RIPE\ NCC}
We use the \acp{URL} of the \ac{RRDP} server as the identifier for each repository.

For the most part of this paper, we focus on the measurements run on December 6 2025.
We validated that our measurement are representative for normal
repository behaviour by performing two additional measurements two days before and two
days after the initial measurement. 

\subsection{Measurements}
\paragraph{Hostnames (I-D Section 5)}\label{sec:hostnames}
The I-D recommends the use of different hostnames for the publication, \ac{RRDP} and rsync server to allow for dedicated 
infrastructure.
For each \ac{CA} certificate, we compare if the \ac{FQDN} of the \ac{URL} of the \ac{RRDP} and the \ac{URL} of the rsync server are different
at different levels of the DNS hierarchy. We compare the difference of the
hostname (left most label, \textbf{Different hostnames}), the 2nd level domain name (the first label left
of the \ac{TLD} - we define the \ac{TLD} using the public suffix list, \textbf{Different subdomains}), and the \ac{TLD}
(\textbf{Different TLDs}).

Also, we verify if the A and AAAA records of the hostnames
 are validly signed using \ac{DNSSEC}
(\textbf{DNSSEC (RRDP)} and \textbf{DNSSEC (rsync)}).

\paragraph{IP Address Space and Autonomous Systems (I-D Section 6)}
If a repository hosts \acp{ROA} that cover the IP address space used for its own \ac{RRDP} and rsync server, 
there is a risk of the repository making itself unreachable, without the ability to publish remediating objects.
The I-D therefore recommends that setups like this should be avoided.
To analyse the dependency of a repositories' reachability on its own \ac{ROA} objects (\textbf{Reachability ind. of ROA in repo}), we loop through the ROAs distributed by the repository and look for a \ac{ROA} that covers the prefix(es) we determined for the service endpoints. If we find a matching ROA, we mark this prefix as being dependent on its own repository. A repository is deemed as independent of its own endpoints if none of the IP address space is covered by a \ac{ROA} in the repository.

Furthermore, the \ac{BCP} recommends hosting the \ac{RRDP} and rsync server in different networks.
We query the A and AAAA for the \acp{URI} of the servers and determine the matching routed prefix for each IP address. 
We do this using the Network Info API of RIPEstat~\cite{ripe-stat-network-info}. Based on this we determine if the \ac{RRDP} and rsync servers 
of a repository are hosted in different networks (\textbf{RRDP and rsync in diff. networks}). We mark this as fulfilled when the repository uses different prefixes for both servers and both IP address families. Having no IP addresses of a family type for one of the servers does not count as a different network.

\paragraph{Same origin URIs (I-D Section 7.1)}\label{sec:same_origin}
The I-D remarks that as defined in RFC9674, the ``uri'' attributes in notification files need to refer to the same \ac{URI} for all listed snapshot and delta files \cite{rfc9674}.
We download the content of each notification file and analyse the \acp{URI} in
it pointing to delta and snapshot files. If they point to the same host as
the notification file \ac{URI}, we deem the \ac{RRDP} server as compliant to
this requirement (\textbf{Same origin}). In addition, we check if a HTTP
redirect occurs when fetching the content.

\paragraph{Endpoint Protection (I-D Section 7.2)}\label{sec:endpoint_protection}
This \ac{BCP} recommends the use of access control to protect \ac{RRDP} endpoints.
Here, we send an HTTP request to each \ac{URL} pointing towards a notification file and record the returned HTTP status
using the following methods: GET, OPTIONS, HEAD, POST, PUT, PATCH, DELETE (\textbf{Endpoint protection}).
The RFCs defining the RRDP protocol do not specify which methods and parameters
are mandatory. We assume that a web server implements endpoint protection if
it only answers with status code 200 for the methods GET and HEAD and with a
status code between 400 and 599 for the methods PATCH and DELETE.\footnote{This is in
accordance with Nmap (\url{https://nmap.org/nsedoc/scripts/http-methods.html}).}

\paragraph{Bandwidth and Data Usage (I-D Section 7.3)}
This practice recommends to support compression to save bandwidth.
Our test consists of downloading each notification file.
If the server serves the notification file in a compressed form
(i.e., when the \emph{Content-Encoding} header is set to something like \emph{gzip} or \emph{deflate}),
we mark the \ac{RRDP} server as compliant with this requirement (\textbf{Compression}).

\paragraph{Content Availability (I-D Section 7.4)}\label{sec:methodology_content_availability}
The I-D's authors recommend to use a \ac{CDN} to achieve high availability.
We use two methods to determine whether a \ac{RRDP} Server relies on a \ac{CDN} to
distribute its content (\textbf{CDN}).

First, we test if the IP addresses of
the \ac{RRDP} Server are announced using \ac{BGP} anycast using a list IP prefixes compiled by LaCeS~\cite{hendriks2025laces}.
We use a list of IP prefixes compiled by LaCeS~\cite{hendriks2025laces}.
Second,
we test whether the IP addresses of the \ac{RRDP} server are announced by
\ac{AS} numbers that are categorized as \emph{CDN} by bgp.tools~\cite{bgp-tools-cdn-tags-2025}.
To find out which \ac{AS} (or \acp{AS}) announces a particular IP address,
we use a data set provided by iptoasn.com~\cite{iptoasn-ip2asn-combined-2025}.

For the most part of the paper, we classify an endpoint as relying on a \ac{CDN} if either of the two methods return a positive result.

\paragraph{Limit Notification File Size (I-D Section 7.5)}\label{sec:method_notification_file_size}
This \ac{BCP} recommends practices to reduce the amount of data a \ac{RP} needs to retrieve.
For each repository, we collect the size of the notification, snapshot and
delta file. We take the sum
of the size of each delta files listed in the corresponding
notification file (\textbf{Snapshot $>$ Deltas}). To save resources of the
\ac{RRDP} server, we attempt to look up the file size indicated in the response
header and stop the download afterwards whenever possible.

Also, we measure how long the \ac{RRDP} server keeps individual delta files (\textbf{Delta older than 4h}) by downloading each notification file every 30 seconds for 24 hours. We record when we have
observed each delta file for the first and for the last time. We use this
information also to measure the
update frequency of the publication server (\textbf{Delta frequency larger than 1min}). Note that we can only identify
update frequencies of 30 seconds or longer due to our measurement frequency.

\paragraph{Manifest and CRL Update Times (I-D Section 7.6)}
The I-D discusses up- and downsides of longer and shorter validity periods for manifests and \acp{CRL}.
It does not give a clear recommendation. To understand which validity times are common, we measure
the average for each repository. We collect the update times from manifests and \acp{CRL} (\textbf{Manifest/CRL validity}). For both, we extract the ``next update'' and ``this update'' fields and report the average difference between the two.

\subsection{Not measured practices and ethical considerations} 

We focus on practices that we can measure externally, without the need of
ingesting content into the RPKI, thus excluding the publication server. Other aspects can
only be observed in case of problems at the CA (e.g.\ I-D Section 4.4 Data
Loss). Also, we do not measure Notification File Timing (I-D Section 7.7.1) and
L4 Load-Balancing (I-D Section 7.7.2) as those require frequent, distributed
measurements with low chance of actually finding mis-behaviour.
Finally, we do not measure the implementation of the rsync server. Some
recommendations are not visible for an external observer (e.g.\ parts of Load
Balancing and Testing, I-D Section 8.3) or require highly frequent
measurements, possibly draining the resources of the server (e.g.\ Consistent
Content, I-D Section 8.1).

In our measurements, we do not put an unreasonable burden on the measured
infrastructure. One measurement consists of 2 runs of the rpki-client.
Overall, we perform 3 measurements spread over 6 days. The collected data is
public and does not contain personal identifiable information.

\section{Results}

\subsection{Overview of repositories}

In this subsection, we provide an overview of the observed
repositories and the implemented \acp{BCP}. In the later subsections, we
discuss in more detail who implements the different \acp{BCP} and how.

We observed 100 unique repositories (identified by \acp{URL} pointing to \ac{RRDP}
notification files). For 10 of these 100 repositories, we were unable to fetch the
notification file for different reasons. 
We do not take these 10 repositories
into account for our analysis. Out of the 90 remaining \acp{URL}, 
26 point to notification files hosted by AWS. These 26 \acp{URL} are used by individual repositories, but they all implement the same
practices. If necessary, we point out the effect of the repositories by AWS
on our results.

\paragraph{ROA payloads}

\begin{figure}
  \centering
  \includegraphics[width=0.68\textwidth]{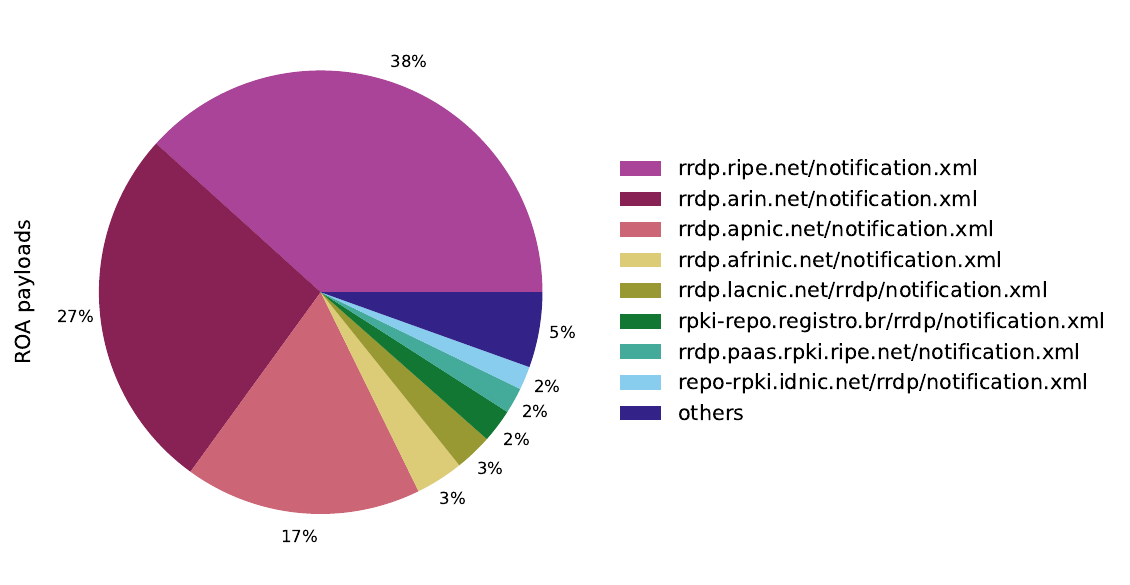}
  \caption{Distribution of ROA payloads rounded to percent}
  \label{fig:roa_payloads}
\end{figure}

\Cref{fig:roa_payloads} shows the distribution of \ac{ROA} payloads. 
The \acp{RIR} take up the first 5 positions and with that 91.12\% of \ac{ROA} payloads distributed through the RPKI. All repositories in the top 8 of biggest repositories are hosted by \acp{RIR} and 
\acp{NIR}. When grouping \ac{AWS}' repositories together, they come in at position 5 with 2.88\% of the total \ac{ROA} payloads.

In comparison, 81.93\% of \ac{ROA} objects (potentially containing more than one payload) are distributed through the servers of the 5 \acp{RIR}. With 5.08\% of ROAs, the biggest repository not hosted by an \ac{RIR} is the of \ac{NIR} registro.br. \Cref{fig:roas} visualising the distribution of \ac{ROA} objects can be found in the appendix.
This shows that even though more than 80 self-hosted repositories exist, the vast majority of Internet address owners still rely on the publication servers and repositories of their parents, following the recommendation in I-D Section 4.1.

\begin{figure}
  \centering
  \begin{minipage}[t]{0.45\linewidth}
    \includegraphics[width=\textwidth]{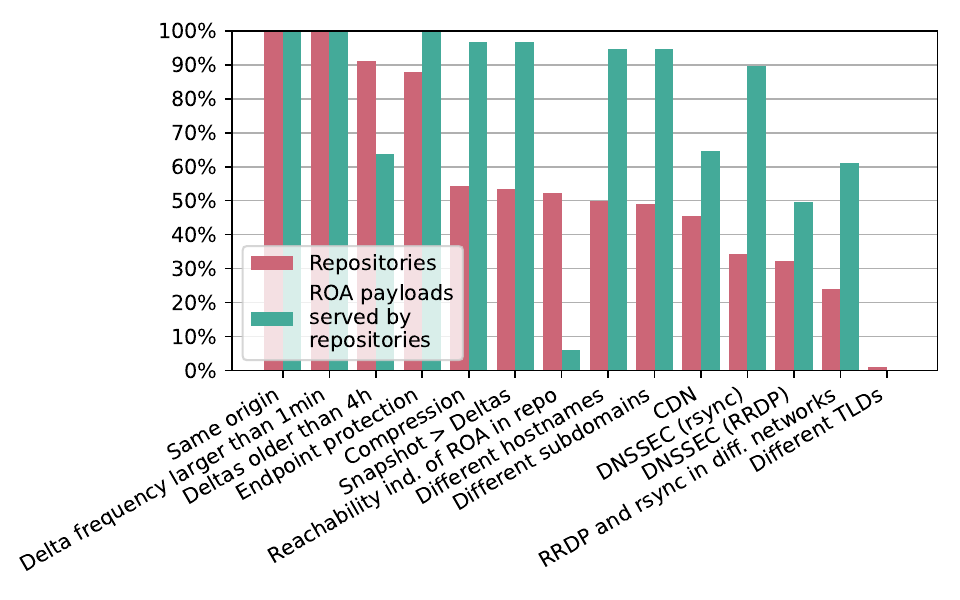}
    \caption{Share of repositories implementing a \ac{BCP} (red bars). ROA payloads served by repositories implementing each \ac{BCP} (green bars).}
    \label{fig:bcps_share}
  \end{minipage}
  \quad
  \begin{minipage}[t]{0.45\linewidth}
    \includegraphics[width=\textwidth]{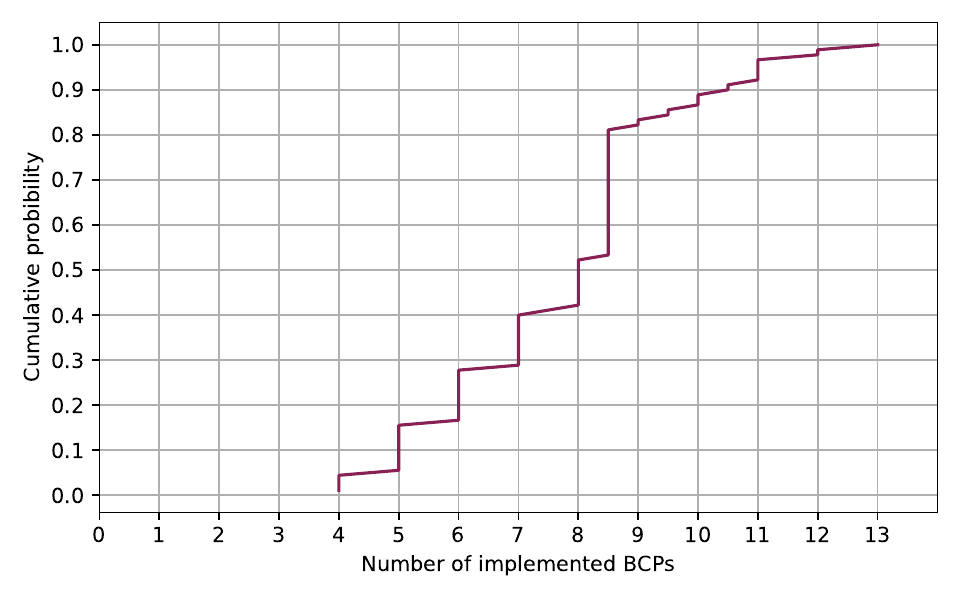}
    \caption{CDF of the number \acp{BCP} each repository implements}
    \label{fig:bcps_per_repo}
  \end{minipage}
\end{figure}

\paragraph{Overview of implemented \acp{BCP}}

\Cref{fig:bcps_share} shows the share of repositories that implement a certain \ac{BCP}.
The only \ac{BCP} that all repositories implement is the \emph{same origin} policy and recommendations regarding the update of delta files. In contrast, only one of the repositories use different \acp{TLD} for their \ac{RRDP} and rsync servers and only 24\% host their \ac{RRDP} and rsync server in different networks.

\Cref{fig:bcps_share} also shows how many \ac{ROA} payloads are served by repositories implementing each \ac{BCP}. While some \acp{BCP} are only implemented by a minority of servers, these servers provide still more than 80\% of all \ac{ROA} payloads in the RPKI. Examples include ``Different Hostnames'', or the recommendation ``Reachability independent of ROA in repository''. %
Overall, we find that repositories responsible for a large number of \ac{ROA} payloads do not necessarily implement more \acp{BCP}: These numbers only weakly positively correlate (Pearson correlation $R = 0.270$).

\paragraph{Individual repositories}

\Cref{fig:bcps_per_repo} shows that 50\% of the repositories implement at least 8 out of the 13 \acp{BCP}. The repositories operated by \ac{AWS} influence this metric (jump at 8.5 \acp{BCP} in \Cref{fig:bcps_per_repo}).\footnote{The servers of AWS \emph{partially} implement \textbf{RRDP and rsync in diff. networks}.} When treating the repositories by \ac{AWS} as one single repository, the median number of implemented \acp{BCP} drops to 7.

The first part of \Cref{table:bcps_rirs_and_nirs} shows how the measured \acp{BCP} are implemented by the \acp{RIR}. 
The second half shows the same, but for \acp{NIR}.
For each repository, the table reports the number of \acp{ROA}, and whether the practices as outlined in~\Cref{sec:methodology} are followed. We especially focus on these repositories since, overall, they host the majority of \acp{ROA}.
A list of the 20 largest repositories with regards to \ac{ROA} count, excluding those hosted by the \acp{RIR} and \acp{NIR}, is part of the appendix (\Cref{table:bcps_20l}). 

\subsection{Hostnames}
45 repositories (50\%) use different hostnames for the RRDP and rsync server. One repository uses different \acp{TLD}. %
\setlength{\tabcolsep}{3.2pt} %
\renewcommand{\arraystretch}{0.9}
\begin{table}[t!]
\raggedright
\footnotesize
\begin{tabular}{lrrrrrrrrrrrrrrr}
 & ROAs & \rot{Different hostnames} & \rot{Different TLDs} & \rot{Different subdomains} & \rot{DNSSEC (RRDP)} & \rot{DNSSEC (rsync)} & \rot{Reachability ind. of ROA in repo} & \rot{RRDP and rsync in diff. networks} & \rot{Same origin} & \rot{Endpoint protection} & \rot{Compression} & \rot{CDN} & \rot{Snapshot > Deltas} & \rot{Delta frequency > 1min} & \rot{Deltas > 4h} \\
\textbf{Repositories by \acp{RIR}} &  &  &  &  &  &  &  &  &  &  &  &  &  &  & \\
rrdp-rps.arin.net/... & 2091 & \cellcolor[HTML]{21918c} \color[HTML]{f1f1f1} \ding{51} & \cellcolor[HTML]{440154} \color[HTML]{f1f1f1} \ding{55} & \cellcolor[HTML]{21918c} \color[HTML]{f1f1f1} \ding{51} & \cellcolor[HTML]{21918c} \color[HTML]{f1f1f1} \ding{51} & \cellcolor[HTML]{21918c} \color[HTML]{f1f1f1} \ding{51} & \cellcolor[HTML]{21918c} \color[HTML]{f1f1f1} \ding{51} & \cellcolor[HTML]{440154} \color[HTML]{f1f1f1} \ding{55} & \cellcolor[HTML]{21918c} \color[HTML]{f1f1f1} \ding{51} & \cellcolor[HTML]{21918c} \color[HTML]{f1f1f1} \ding{51} & \cellcolor[HTML]{21918c} \color[HTML]{f1f1f1} \ding{51} & \cellcolor[HTML]{440154} \color[HTML]{f1f1f1} \ding{55} & \cellcolor[HTML]{21918c} \color[HTML]{f1f1f1} \ding{51} & \cellcolor[HTML]{21918c} \color[HTML]{f1f1f1} \ding{51} & \cellcolor[HTML]{440154} \color[HTML]{f1f1f1} \ding{55} \\
rrdp.arin.net/... & 183288 & \cellcolor[HTML]{21918c} \color[HTML]{f1f1f1} \ding{51} & \cellcolor[HTML]{440154} \color[HTML]{f1f1f1} \ding{55} & \cellcolor[HTML]{21918c} \color[HTML]{f1f1f1} \ding{51} & \cellcolor[HTML]{21918c} \color[HTML]{f1f1f1} \ding{51} & \cellcolor[HTML]{21918c} \color[HTML]{f1f1f1} \ding{51} & \cellcolor[HTML]{440154} \color[HTML]{f1f1f1} \ding{55} & \cellcolor[HTML]{440154} \color[HTML]{f1f1f1} \ding{55} & \cellcolor[HTML]{21918c} \color[HTML]{f1f1f1} \ding{51} & \cellcolor[HTML]{21918c} \color[HTML]{f1f1f1} \ding{51} & \cellcolor[HTML]{21918c} \color[HTML]{f1f1f1} \ding{51} & \cellcolor[HTML]{440154} \color[HTML]{f1f1f1} \ding{55} & \cellcolor[HTML]{21918c} \color[HTML]{f1f1f1} \ding{51} & \cellcolor[HTML]{21918c} \color[HTML]{f1f1f1} \ding{51} & \cellcolor[HTML]{440154} \color[HTML]{f1f1f1} \ding{55} \\
rrdp.apnic.net/... & 16677 & \cellcolor[HTML]{21918c} \color[HTML]{f1f1f1} \ding{51} & \cellcolor[HTML]{440154} \color[HTML]{f1f1f1} \ding{55} & \cellcolor[HTML]{21918c} \color[HTML]{f1f1f1} \ding{51} & \cellcolor[HTML]{21918c} \color[HTML]{f1f1f1} \ding{51} & \cellcolor[HTML]{21918c} \color[HTML]{f1f1f1} \ding{51} & \cellcolor[HTML]{440154} \color[HTML]{f1f1f1} \ding{55} & \cellcolor[HTML]{21918c} \color[HTML]{f1f1f1} \ding{51} & \cellcolor[HTML]{21918c} \color[HTML]{f1f1f1} \ding{51} & \cellcolor[HTML]{21918c} \color[HTML]{f1f1f1} \ding{51} & \cellcolor[HTML]{21918c} \color[HTML]{f1f1f1} \ding{51} & \cellcolor[HTML]{21918c} \color[HTML]{f1f1f1} \ding{51} & \cellcolor[HTML]{21918c} \color[HTML]{f1f1f1} \ding{51} & \cellcolor[HTML]{21918c} \color[HTML]{f1f1f1} \ding{51} & \cellcolor[HTML]{21918c} \color[HTML]{f1f1f1} \ding{51} \\
rpki.rand.apnic.net/rrdp/... & 6 & \cellcolor[HTML]{440154} \color[HTML]{f1f1f1} \ding{55} & \cellcolor[HTML]{440154} \color[HTML]{f1f1f1} \ding{55} & \cellcolor[HTML]{440154} \color[HTML]{f1f1f1} \ding{55} & \cellcolor[HTML]{440154} \color[HTML]{f1f1f1} \ding{55} & \cellcolor[HTML]{440154} \color[HTML]{f1f1f1} \ding{55} & \cellcolor[HTML]{21918c} \color[HTML]{f1f1f1} \ding{51} & \cellcolor[HTML]{440154} \color[HTML]{f1f1f1} \ding{55} & \cellcolor[HTML]{21918c} \color[HTML]{f1f1f1} \ding{51} & \cellcolor[HTML]{21918c} \color[HTML]{f1f1f1} \ding{51} & \cellcolor[HTML]{440154} \color[HTML]{f1f1f1} \ding{55} & \cellcolor[HTML]{440154} \color[HTML]{f1f1f1} \ding{55} & \cellcolor[HTML]{21918c} \color[HTML]{f1f1f1} \ding{51} & \cellcolor[HTML]{21918c} \color[HTML]{f1f1f1} \ding{51} & \cellcolor[HTML]{21918c} \color[HTML]{f1f1f1} \ding{51} \\
rrdp.sub.apnic.net/... & 196 & \cellcolor[HTML]{21918c} \color[HTML]{f1f1f1} \ding{51} & \cellcolor[HTML]{440154} \color[HTML]{f1f1f1} \ding{55} & \cellcolor[HTML]{21918c} \color[HTML]{f1f1f1} \ding{51} & \cellcolor[HTML]{21918c} \color[HTML]{f1f1f1} \ding{51} & \cellcolor[HTML]{21918c} \color[HTML]{f1f1f1} \ding{51} & \cellcolor[HTML]{21918c} \color[HTML]{f1f1f1} \ding{51} & \cellcolor[HTML]{21918c} \color[HTML]{f1f1f1} \ding{51} & \cellcolor[HTML]{21918c} \color[HTML]{f1f1f1} \ding{51} & \cellcolor[HTML]{21918c} \color[HTML]{f1f1f1} \ding{51} & \cellcolor[HTML]{21918c} \color[HTML]{f1f1f1} \ding{51} & \cellcolor[HTML]{21918c} \color[HTML]{f1f1f1} \ding{51} & \cellcolor[HTML]{21918c} \color[HTML]{f1f1f1} \ding{51} & \cellcolor[HTML]{21918c} \color[HTML]{f1f1f1} \ding{51} & \cellcolor[HTML]{21918c} \color[HTML]{f1f1f1} \ding{51} \\
rrdp.afrinic.net/... & 10975 & \cellcolor[HTML]{21918c} \color[HTML]{f1f1f1} \ding{51} & \cellcolor[HTML]{440154} \color[HTML]{f1f1f1} \ding{55} & \cellcolor[HTML]{21918c} \color[HTML]{f1f1f1} \ding{51} & \cellcolor[HTML]{21918c} \color[HTML]{f1f1f1} \ding{51} & \cellcolor[HTML]{21918c} \color[HTML]{f1f1f1} \ding{51} & \cellcolor[HTML]{440154} \color[HTML]{f1f1f1} \ding{55} & \cellcolor[HTML]{21918c} \color[HTML]{f1f1f1} \ding{51} & \cellcolor[HTML]{21918c} \color[HTML]{f1f1f1} \ding{51} & \cellcolor[HTML]{21918c} \color[HTML]{f1f1f1} \ding{51} & \cellcolor[HTML]{21918c} \color[HTML]{f1f1f1} \ding{51} & \cellcolor[HTML]{21918c} \color[HTML]{f1f1f1} \ding{51} & \cellcolor[HTML]{21918c} \color[HTML]{f1f1f1} \ding{51} & \cellcolor[HTML]{21918c} \color[HTML]{f1f1f1} \ding{51} & \cellcolor[HTML]{21918c} \color[HTML]{f1f1f1} \ding{51} \\
rrdp.lacnic.net/rrdp/... & 15757 & \cellcolor[HTML]{21918c} \color[HTML]{f1f1f1} \ding{51} & \cellcolor[HTML]{440154} \color[HTML]{f1f1f1} \ding{55} & \cellcolor[HTML]{21918c} \color[HTML]{f1f1f1} \ding{51} & \cellcolor[HTML]{440154} \color[HTML]{f1f1f1} \ding{55} & \cellcolor[HTML]{440154} \color[HTML]{f1f1f1} \ding{55} & \cellcolor[HTML]{21918c} \color[HTML]{f1f1f1} \ding{51} & \cellcolor[HTML]{440154} \color[HTML]{f1f1f1} \ding{55} & \cellcolor[HTML]{21918c} \color[HTML]{f1f1f1} \ding{51} & \cellcolor[HTML]{21918c} \color[HTML]{f1f1f1} \ding{51} & \cellcolor[HTML]{440154} \color[HTML]{f1f1f1} \ding{55} & \cellcolor[HTML]{440154} \color[HTML]{f1f1f1} \ding{55} & \cellcolor[HTML]{21918c} \color[HTML]{f1f1f1} \ding{51} & \cellcolor[HTML]{21918c} \color[HTML]{f1f1f1} \ding{51} & \cellcolor[HTML]{440154} \color[HTML]{f1f1f1} \ding{55} \\
rrdp.paas.rpki.ripe.net/... & 6583 & \cellcolor[HTML]{21918c} \color[HTML]{f1f1f1} \ding{51} & \cellcolor[HTML]{440154} \color[HTML]{f1f1f1} \ding{55} & \cellcolor[HTML]{21918c} \color[HTML]{f1f1f1} \ding{51} & \cellcolor[HTML]{440154} \color[HTML]{f1f1f1} \ding{55} & \cellcolor[HTML]{21918c} \color[HTML]{f1f1f1} \ding{51} & \cellcolor[HTML]{21918c} \color[HTML]{f1f1f1} \ding{51} & \cellcolor[HTML]{21918c} \color[HTML]{f1f1f1} \ding{51} & \cellcolor[HTML]{21918c} \color[HTML]{f1f1f1} \ding{51} & \cellcolor[HTML]{21918c} \color[HTML]{f1f1f1} \ding{51} & \cellcolor[HTML]{21918c} \color[HTML]{f1f1f1} \ding{51} & \cellcolor[HTML]{21918c} \color[HTML]{f1f1f1} \ding{51} & \cellcolor[HTML]{21918c} \color[HTML]{f1f1f1} \ding{51} & \cellcolor[HTML]{21918c} \color[HTML]{f1f1f1} \ding{51} & \cellcolor[HTML]{440154} \color[HTML]{f1f1f1} \ding{55} \\
rrdp.ripe.net/... & 45338 & \cellcolor[HTML]{21918c} \color[HTML]{f1f1f1} \ding{51} & \cellcolor[HTML]{440154} \color[HTML]{f1f1f1} \ding{55} & \cellcolor[HTML]{21918c} \color[HTML]{f1f1f1} \ding{51} & \cellcolor[HTML]{440154} \color[HTML]{f1f1f1} \ding{55} & \cellcolor[HTML]{21918c} \color[HTML]{f1f1f1} \ding{51} & \cellcolor[HTML]{440154} \color[HTML]{f1f1f1} \ding{55} & \cellcolor[HTML]{21918c} \color[HTML]{f1f1f1} \ding{51} & \cellcolor[HTML]{21918c} \color[HTML]{f1f1f1} \ding{51} & \cellcolor[HTML]{21918c} \color[HTML]{f1f1f1} \ding{51} & \cellcolor[HTML]{21918c} \color[HTML]{f1f1f1} \ding{51} & \cellcolor[HTML]{21918c} \color[HTML]{f1f1f1} \ding{51} & \cellcolor[HTML]{21918c} \color[HTML]{f1f1f1} \ding{51} & \cellcolor[HTML]{21918c} \color[HTML]{f1f1f1} \ding{51} & \cellcolor[HTML]{21918c} \color[HTML]{f1f1f1} \ding{51} \\
\textbf{Repositories by \acp{NIR}} &  &  &  &  &  &  &  &  &  &  &  &  &  \\
rpki-repo.registro.br/rrdp/... & 17421 & \cellcolor[HTML]{440154} \color[HTML]{f1f1f1} \ding{55} & \cellcolor[HTML]{440154} \color[HTML]{f1f1f1} \ding{55} & \cellcolor[HTML]{440154} \color[HTML]{f1f1f1} \ding{55} & \cellcolor[HTML]{440154} \color[HTML]{f1f1f1} \ding{55} & \cellcolor[HTML]{440154} \color[HTML]{f1f1f1} \ding{55} & \cellcolor[HTML]{440154} \color[HTML]{f1f1f1} \ding{55} & \cellcolor[HTML]{440154} \color[HTML]{f1f1f1} \ding{55} & \cellcolor[HTML]{21918c} \color[HTML]{f1f1f1} \ding{51} & \cellcolor[HTML]{21918c} \color[HTML]{f1f1f1} \ding{51} & \cellcolor[HTML]{21918c} \color[HTML]{f1f1f1} \ding{51} & \cellcolor[HTML]{440154} \color[HTML]{f1f1f1} \ding{55} & \cellcolor[HTML]{21918c} \color[HTML]{f1f1f1} \ding{51} & \cellcolor[HTML]{21918c} \color[HTML]{f1f1f1} \ding{51} & \cellcolor[HTML]{440154} \color[HTML]{f1f1f1} \ding{55} \\
repo-rpki.idnic.net/rrdp/... & 8464 & \cellcolor[HTML]{440154} \color[HTML]{f1f1f1} \ding{55} & \cellcolor[HTML]{440154} \color[HTML]{f1f1f1} \ding{55} & \cellcolor[HTML]{440154} \color[HTML]{f1f1f1} \ding{55} & \cellcolor[HTML]{440154} \color[HTML]{f1f1f1} \ding{55} & \cellcolor[HTML]{440154} \color[HTML]{f1f1f1} \ding{55} & \cellcolor[HTML]{440154} \color[HTML]{f1f1f1} \ding{55} & \cellcolor[HTML]{440154} \color[HTML]{f1f1f1} \ding{55} & \cellcolor[HTML]{21918c} \color[HTML]{f1f1f1} \ding{51} & \cellcolor[HTML]{21918c} \color[HTML]{f1f1f1} \ding{51} & \cellcolor[HTML]{21918c} \color[HTML]{f1f1f1} \ding{51} & \cellcolor[HTML]{440154} \color[HTML]{f1f1f1} \ding{55} & \cellcolor[HTML]{21918c} \color[HTML]{f1f1f1} \ding{51} & \cellcolor[HTML]{21918c} \color[HTML]{f1f1f1} \ding{51} & \cellcolor[HTML]{440154} \color[HTML]{f1f1f1} \ding{55} \\
rpki-repository.nic.ad.jp/rrdp/ap/... & 5533 & \cellcolor[HTML]{440154} \color[HTML]{f1f1f1} \ding{55} & \cellcolor[HTML]{440154} \color[HTML]{f1f1f1} \ding{55} & \cellcolor[HTML]{440154} \color[HTML]{f1f1f1} \ding{55} & \cellcolor[HTML]{21918c} \color[HTML]{f1f1f1} \ding{51} & \cellcolor[HTML]{21918c} \color[HTML]{f1f1f1} \ding{51} & \cellcolor[HTML]{440154} \color[HTML]{f1f1f1} \ding{55} & \cellcolor[HTML]{440154} \color[HTML]{f1f1f1} \ding{55} & \cellcolor[HTML]{21918c} \color[HTML]{f1f1f1} \ding{51} & \cellcolor[HTML]{21918c} \color[HTML]{f1f1f1} \ding{51} & \cellcolor[HTML]{21918c} \color[HTML]{f1f1f1} \ding{51} & \cellcolor[HTML]{440154} \color[HTML]{f1f1f1} \ding{55} & \cellcolor[HTML]{21918c} \color[HTML]{f1f1f1} \ding{51} & \cellcolor[HTML]{21918c} \color[HTML]{f1f1f1} \ding{51} & \cellcolor[HTML]{21918c} \color[HTML]{f1f1f1} \ding{51} \\
rrdp.twnic.tw/rrdp/... & 3929 & \cellcolor[HTML]{21918c} \color[HTML]{f1f1f1} \ding{51} & \cellcolor[HTML]{440154} \color[HTML]{f1f1f1} \ding{55} & \cellcolor[HTML]{21918c} \color[HTML]{f1f1f1} \ding{51} & \cellcolor[HTML]{21918c} \color[HTML]{f1f1f1} \ding{51} & \cellcolor[HTML]{21918c} \color[HTML]{f1f1f1} \ding{51} & \cellcolor[HTML]{440154} \color[HTML]{f1f1f1} \ding{55} & \cellcolor[HTML]{440154} \color[HTML]{f1f1f1} \ding{55} & \cellcolor[HTML]{21918c} \color[HTML]{f1f1f1} \ding{51} & \cellcolor[HTML]{21918c} \color[HTML]{f1f1f1} \ding{51} & \cellcolor[HTML]{440154} \color[HTML]{f1f1f1} \ding{55} & \cellcolor[HTML]{21918c} \color[HTML]{f1f1f1} \ding{51} & \cellcolor[HTML]{21918c} \color[HTML]{f1f1f1} \ding{51} & \cellcolor[HTML]{21918c} \color[HTML]{f1f1f1} \ding{51} & \cellcolor[HTML]{21918c} \color[HTML]{f1f1f1} \ding{51} \\
rpki.cnnic.cn/rrdp/... & 2047 & \cellcolor[HTML]{440154} \color[HTML]{f1f1f1} \ding{55} & \cellcolor[HTML]{440154} \color[HTML]{f1f1f1} \ding{55} & \cellcolor[HTML]{440154} \color[HTML]{f1f1f1} \ding{55} & \cellcolor[HTML]{440154} \color[HTML]{f1f1f1} \ding{55} & \cellcolor[HTML]{440154} \color[HTML]{f1f1f1} \ding{55} & \cellcolor[HTML]{21918c} \color[HTML]{f1f1f1} \ding{51} & \cellcolor[HTML]{440154} \color[HTML]{f1f1f1} \ding{55} & \cellcolor[HTML]{21918c} \color[HTML]{f1f1f1} \ding{51} & \cellcolor[HTML]{21918c} \color[HTML]{f1f1f1} \ding{51} & \cellcolor[HTML]{21918c} \color[HTML]{f1f1f1} \ding{51} & \cellcolor[HTML]{440154} \color[HTML]{f1f1f1} \ding{55} & \cellcolor[HTML]{440154} \color[HTML]{f1f1f1} \ding{55} & \cellcolor[HTML]{21918c} \color[HTML]{f1f1f1} \ding{51} & \cellcolor[HTML]{21918c} \color[HTML]{f1f1f1} \ding{51} \\
\end{tabular}
\caption{Implemented BCPs by repositories operated by \acp{RIR} and \acp{NIR}}
\label{table:bcps_rirs_and_nirs}
\end{table}

8 out of 9 \ac{RIR} hosted repositories use different hostnames for their \ac{RRDP} and rsync servers. Only the \emph{rpki.rand.apnic.net} repository does not comply with this recommendation. Within the group of \ac{NIR} repositories, only 1 repository makes use of different hostnames.
One other noteworthy repository is \emph{magellan.ipxo.com} as it is the only repository that uses a subdomain of its \ac{RRDP} server hostname for its rsync server.

22 (24\%) repositories sign their A and AAAA records for both, their \ac{RRDP} 
server and their rsync server. These servers are responsible for
52\% (\ac{RRDP}) and 92\% (rsync) of all \ac{ROA} payloads, including 5 out
of 9 servers operated by the \acp{RIR} (see \Cref{table:bcps_rirs_and_nirs}). 
The large difference between \ac{RRDP} and rsync
is caused by the RIPE repositories. RIPE NCC hosts the zone of the
rsync server itself, while Akamai hosts the
zone of the RRDP server. The RIPE NCC confirmed us that in their specific
setup, Akamai does not support \ac{DNSSEC}. %

\subsection{IP address space and autonomous systems}

\paragraph{IPv4 and IPv6}

Repositories use a mean of 3.24 IPv4 addresses and a mean of 3.14 IPv6 addresses to host their repository (rsync and \ac{RRDP} combined).
The median for both address families is 1. There is 1 repository that makes use of unreachable IP addresses for its endpoints.

On average more IP addresses are used for the \ac{RRDP} server compared to the rsync server, 2.57 and 1.08 respectively. When only looking at IPv6 addresses, this difference 
is even bigger: on average, 3.01 IPv6 addresses are used for \ac{RRDP} servers and only 0.5 for rsync servers.

\paragraph{Endpoints' reachability dependent on \acp{ROA} in repository}
40 repositories (44\%) distribute at least a subset of the \acp{ROA} covering the IP ranges used for their service endpoints through their own repository. The remaining repositories either distribute \acp{ROA} for the used IP ranges through a different repository or do not publish \acp{ROA} for them at all.

\paragraph{Use of different networks}
37 repositories (41\%) make use of different networks for their \ac{RRDP} and rsync endpoints for at least one IP address family. 6 repositories use different networks for both IPv4 and IPv6, 5 of those are repositories hosted by \acp{RIR}.

\subsection{RRDP server}\label{subsec:rrdp}

\paragraph{Same origin URIs}
All repositories host the delta and snapshots files under the same hostname as the notification file and are therefore compliant with the same origin recommendation. One repository makes use of an HTTP redirect when content is being fetched. However, this repository does not host any \acp{ROA}.

\paragraph{Endpoint Protection}
79 repositories (88\%) deploy endpoint protection.
For the RRDP servers included in~\Cref{table:bcps_rirs_and_nirs} and in the largest 20 remaining servers, there are only 2 servers
that do not deploy endpoint protection.
That means that most of the servers failing our test deploy a small amount of \acp{ROA}.

\paragraph{Bandwidth and Data Usage}\label{sec:res_bandwidth}
55 repositories (61\%) support HTTP compression.
2 repositories operated by \acp{RIR} do not support compression,
but one of them hosts just 6 \acp{ROA}.
With the \ac{NIR}-operated repositories,
1 repository does not support compression.
5 out of the 20 largest remaining servers do not support compression.
As such,
most of the repositories that do not support compression reside in the ``long tail'' of smaller repositories.

\paragraph{Content Availability}
41 repositories (46\%) use a \ac{CDN}.
Among the \acp{RIR},
ARIN does not use a \ac{CDN}.
There are 3 repositories listed for APNIC's,
2 of which use a \ac{CDN}, and
one of them does not.
However, the latter hosts just 6 \acp{ROA}.
Furthermore,
LACNIC's server also does not use a CDN.
Among the \acp{NIR},
we notice that all but one of the \acp{NIR} do \emph{not} use a CDN to distribute their data.
Among the other repositories,
all repositories that have more than 1,000 \acp{ROA} use a CDN.

\paragraph{Notification file size}\label{sec:res_notification_file_size}

Since the content of the repository can change, we requested their contents 3
times over the course of 6 days.
81 (90\%) RRDP servers never serve snapshot files exceeding 15,516~KB. The
total size of delta files never exceeds 13.554~KB for 90\% of the RRDP servers. See for more details \Cref{fig:snapshot_delta_cdf} in the appendix.
The RRDP server of ARIN serve the largest snapshot file (623,152~KB),
while a server of \ac{AWS} serves the largest total size of delta files (144,535~KB). 
For the majority of the repositories, the size of the snapshot file stays
 stable over the course of our measurements (standard
deviation $phi < 26~KB$ for
90\% of all repositories). However, the total size of the delta files varies
stronger (for 17\% of the repositories $phi > 100~KB$). This
is mostly the case for repositories with presumably high activity, like the
ones by the \acp{RIR}.

\begin{figure}
  \begin{minipage}[t]{0.38\linewidth}  
    \includegraphics[width=\textwidth]{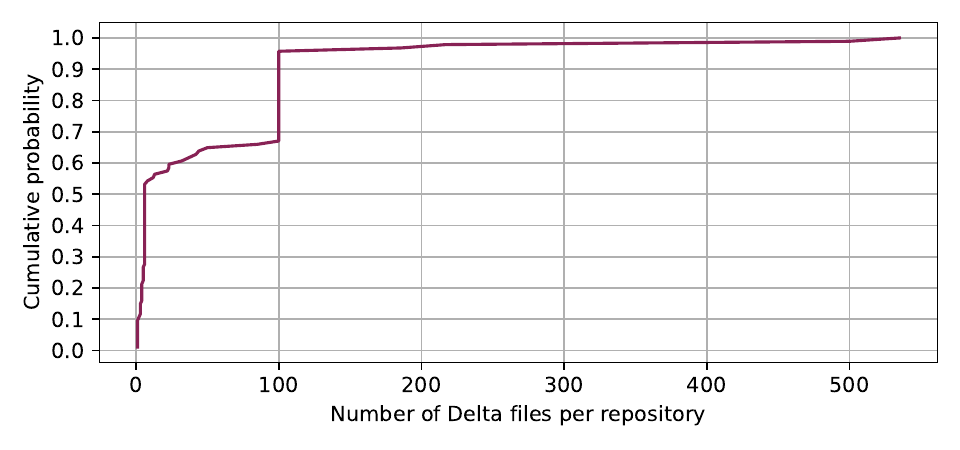}
    \caption{Distribution of number of delta files per repository}%
    \label{fig:delta_files}
  \end{minipage}
  \quad
  \begin{minipage}[t]{0.38\linewidth}  
    \includegraphics[width=\textwidth]{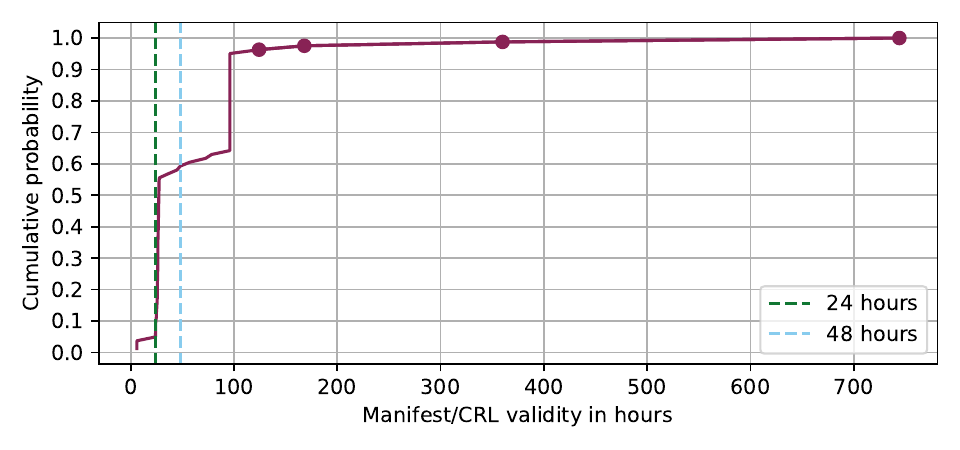}
    \caption{Distribution of average manifest/CRL validity per repository marking the ones with > 4 days}
    \label{fig:manifest_validity}
  \end{minipage}
\end{figure}

For 44 (49\%) repositories, the sum of delta files is always larger than the
snapshot file (5 times larger on average). These repositories thus do not follow the
recommendation. Again, the repositories by \ac{AWS} stand out, with the total size
of delta files exceeding the size of the snapshot files between 4 and 104
times. Instead, the \ac{RRDP} servers by \ac{AWS}
 handle a maximum number of 100 deltas files (see jump at 100 at \Cref{fig:delta_files}). Also the
main repository by APNIC and the repository by the \ac{NIR} registro.br have
enforced a limit of 500 and 50 delta files respectively.

74 (82\%) of all repositories keep delta files for at least 4 hours, thus follow
the recommended practice. 6 never follow this recommendation and 10 do
not show consistent behaviour. 58\% of all repositories always keep their delta
files for at least 1 day (the maximum time period we measured). %
Finally, 83 (92\%) of the measured repositories always publish new delta files less
frequently than once every minute.

\paragraph{Manifest/CRL validity}
82 repositories (91\%) publish minimum one valid manifest or CRL. The remaining repositories do not host any manifests and \acp{CRL} or none with a valid signature chain.

\Cref{fig:manifest_validity} shows the distribution of the average validity time per repository. Here, we only include repositories with minimum one valid manifest or CRL. Because the average manifest validity does not differ significantly from the average \ac{CRL} validity for any of the repositories,\footnote{There are 2 repositories where it does differ, however, by less than 2 milliseconds.} we discuss them together. The I-D describes that ``one large repository'' decided to adjust their reissunance cycle from once every 24 hours to once every 48 hours. To illustrate where most repositories fall with regards to that, the 24 and 48 hour timestamps are marked in \Cref{fig:manifest_validity}.

Most repositories have an average manifest and \ac{CRL} validity time between 1 and 2 days (median 1 day and 03:15:50). When rounding the validity period to days, 1 day is the most common value with 43 repositories (48\%) of the evaluated repositories, followed by 4 days with 26 repositories (27\%). These 26 are all hosted by \ac{AWS}. 
The minimum average validity is 6 hours (3 repositories), the maximum average validity is 31 days 00:05:00 (1 repository). 
Only 4 repositories (4\%) have a average validity longer than 4 days, they are marked in \Cref{fig:manifest_validity}.

Rounded to days, RIPE's main repository has an average validity period of 1 day. ARIN's and AFRINIC's main repositories have an average of around 2 days. For LACNIC and APNIC we see longer average validity periods of 5 and 7 days respectively.

\section{Discussion}

\subsection{Common ground and uncommon practices}

There is some common ground on how to operate a repository,
even before the best practices document is published. From \Cref{table:bcps_rirs_and_nirs}, 
we can derive that serving the content from the
same origin and enabling some basic form of endpoint protection, 
are two configuration knobs most operators can agree on. 

In contrast, some suggested practices are rarely
implemented. For example, almost none of the operators choose for different \acp{TLD} for
the rsync and \ac{RRDP} servers and also many domain names of \ac{RRDP} endpoints
are not DNSSEC-signed. 

Some practices are mostly implemented by repositories that host many \ac{ROA}
payloads, for example relying on different hostnames and distributing \ac{RRDP} servers using \acp{CDN}. 
While the use of a \ac{CDN} might be perceived as unnecessary for many
small repositories, the security of their network resources depends on
the availability of their repositories. Putting effort into highly
available \ac{RRDP} servers should therefore not be underestimated.

\subsection{Potentially problematic practices}

We observed some practices that should be adjusted because they may
cause unnecessary burden for \acp{RP} or put the availability of the
repository at an unnecessary risk.

RRDP servers providing larger snapshot and delta files and their
clients could benefit the most from HTTP compression. However, as discussed
in \Cref{sec:res_bandwidth}, some RRDP servers do not support compression and
we find that the median snapshot and delta file size that these servers serve
is higher than of the servers that do support compression.

Also, the I-D does not give a clear recommendation on manifest/CRL validity but remarks that longer and shorter durations have certain up- and downsides.

One downside of short validity periods is that \acp{ROA} might become invalid while a repository is unreachable. 
Of the 3 repositories with an average validity time of less than 1 day (6 hours), 2 distribute less than 10 \ac{ROA} payloads.
One however, \emph{rpki.ccnic.cn}, distributes over 2000 ROA payloads and the short validity time could lead to \acp{ROA} becoming invalid rather quickly if the repository becomes unavailable for example due to unexpected operational issues.

The high average validity periods of 15 and ~31 days that we measured for 2 repositories\footnote{Seen for \emph{rpki-repository.nic.ad.jp} and  \emph{akane.maru.co.jp} respectively.} makes these more vulnerable to potential replay attacks with old \ac{ROA} objects \cite{krill_docs}.

\subsection{Alternatives to \acp{BCP}}
\label{sec:discussion_alternative_approaches}

As also noted in the I-D, there might be good reasons to diverge from the suggested practices.  

For example, 
not using a \ac{CDN} does not mean a service cannot have a good level of availability.
ARIN's server, for instance, is classified as not using a \ac{CDN}.
However,
the hostname of that server (\emph{rrdp.arin.net}) resolves to 6 IPv4 addresses and 6 IPv6 addresses.
For both address families, 3 of the 6 IP addresses are announced by \ac{AS} 394018,
and the other 3 are announced by \ac{AS} 393225,
also connected at different \acp{IXP}.
As such,
in terms of geographical location and connectivity to the Internet,
the 2 groups of servers serving \emph{rrdp.arin.net} seem somewhat independent from each other.

Also, there might be good arguments for not signing the \ac{DNS} records of
the publication servers with \ac{DNSSEC}. \ac{DNS} spoofing would only have limited effect since the integrity and 
 authenticity of the information in RPKI itself is cryptographically protected. 

\subsection{Measurement limitations}
We faced several limitations. First, there is not a definite way of measuring compliance with \textbf{Endpoint protection} and the I-D does not provide guidelines.
We picked an interpretation of this recommendation that can be measured (see~\Cref{sec:endpoint_protection}), but acknowledge that the test could cover more scenarios. Failing our test does not mean that the configuration of the server is wrong.

Also, while we did observe that a substantial amount of repositories make use of different hostnames for their \ac{RRDP} and rsync servers, those hostnames could still rely on the same underlying \ac{DNS} infrastructure (\textbf{Different hostnames}).

Next, the use of different IP addresses does not necessarily tell us about the redundancy and independency of the underlying infrastructure (\textbf{RRDP and rsync in diff. networks}). The different IPs could still lead to the same physical infrastructure.

Finally, some repositories might make use of different networks for their endpoints, where some are covered by a \ac{ROA} in different repositories (\textbf{Reachability ind. of ROA in repo}). We did not measure if there are any \acp{ROA} covering (part of) the IP address space in other repositories. These could be used to distribute remediating updates in case a \ac{ROA} in the repository makes its own origination invalid and with that becomes temporarily unreachable.

\section{Related Work}

The performance of \ac{RPKI} publication servers has been subject of study before.
Hlavacek et al. \cite{10.1145/3548606.3560645} performed a measurement study into the resilience of \acp{RP},
identifying \ac{DNS} as a weak link in the \ac{RPKI} system.
In another study, Hlavacek et al. \cite{10.1145/3603269.3604861} investigated ways in which \acp{RP} software can be attacked by malicious publication points.
Our work complements Hlavacek et al.'s work because we look at \ac{RPKI} repositories as opposed to \ac{RP} software.

Huston \cite{huston_rpki_resilience} discusses the resilience of the \ac{RPKI}, but does not contain measurements.
Huston concludes that \ac{RPKI}'s resilience is at an acceptable level of resilience at the moment, but needs to be re-evaluated if \ac{RPKI} were to get more or new roles.

Two policy documents at RIPE (published, not yet implemented~\cite{ripe-847})
and APNIC (waiting for endorsement by the Executive Council~\cite{apnic-prop-166}) 
propose the ``revocation of persistently non-functional
\ac{RPKI} \acp{CA}''. We also encountered \acp{CA} of which the \ac{RRDP} and rsync servers were unreachable or of 
which its content was not cryptographically
valid, but in general we focus on the behaviour of the functional CAs and 
their publication servers.

Recently, Schulmann et al. \cite{10.1145/3719027.3760715} proposed a method to
fingerprint \ac{RPKI} repositories. While not the goal of our research, some of
the implementation details studied in this paper could be useful for
fingerprinting repositories (e.g.\ the chosen maximum number of delta files
or the update frequency). 

Currently, a new protocol called Erik is
being developed in the \ac{IETF}~\cite{ietf-sidrops-rpki-erik-protocol-01}. 
Erik adds an additional distribution layer which could improve the availability of information in
the \ac{RPKI}, rendering some of the recommended practices in the I-D \cite{ietf-sidrops-publication-server-bcp-05} less urgent.

\section{Conclusion}

Even before the ``RPKI Publication Server Best Current
Practices'' I-D has become an official RFC, the major share of objects in
the \ac{RPKI} are already published by servers implementing many of the
suggested best practices. At the same time, there is still plenty room for
improvement, e.g.\ when increasing resilience against attacks and outages in
the \ac{DNS}. Lastly, there are operational practices that diverge
significantly from the suggested best practices (e.g. the total size of the
published delta files). Future work could study the motivation of these
design choices. As soon as the RFC is published, measuring the uptake of
the suggested best practices at publication servers should be considered.

\label{lastpage}

\bibliographystyle{ACM-Reference-Format.bst}
\balance
\bibliography{library}

\newpage
\begin{appendices}
\crefalias{section}{appendix}\label{app:figures}

\begin{figure}[!h]
  \centering
  \includegraphics[width=0.7\textwidth]{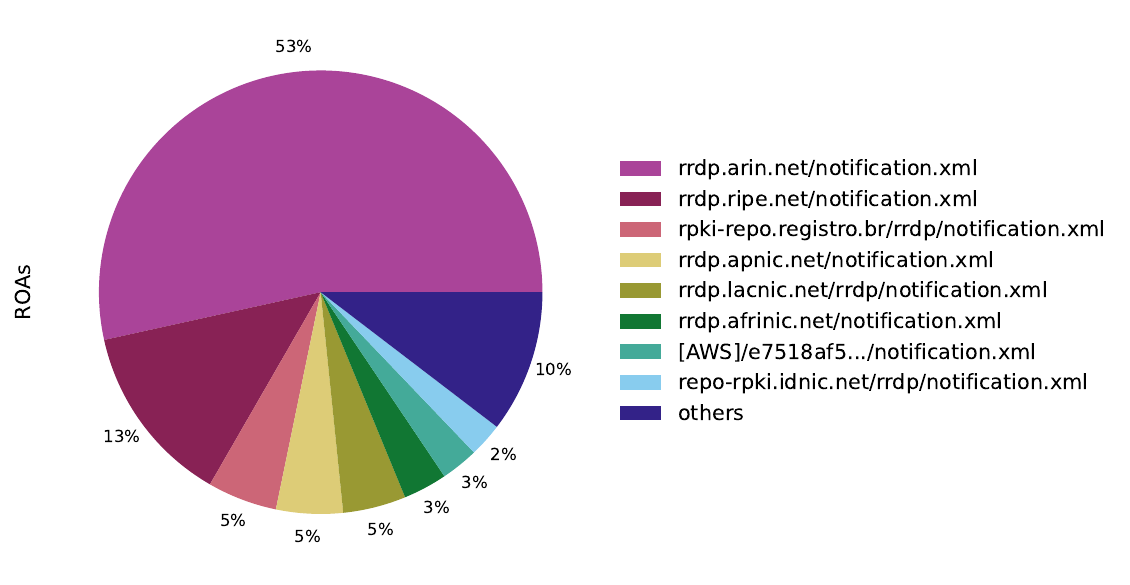}
  \caption{Distribution of \acp{ROA} rounded to percent}
  \label{fig:roas}
\end{figure}

\setlength{\tabcolsep}{3.5pt} %
\renewcommand{\arraystretch}{0.9}
\begin{table}[!h]
\raggedright
\footnotesize
\begin{tabular}{lrrrrrrrrrrrrrrr}
 & ROAs & \rot{Different hostnames} & \rot{Different TLDs} & \rot{Different subdomains} & \rot{DNSSEC (RRDP)} & \rot{DNSSEC (rsync)} & \rot{Reachability ind. of ROA in repo} & \rot{RRDP and rsync in diff. networks} & \rot{Same origin} & \rot{Endpoint protection} & \rot{Compression} & \rot{CDN} & \rot{Snapshot > Deltas} & \rot{Delta frequency > 1min} & \rot{Deltas > 4h} \\
{[AWS]}/e7518af5.../... & 9352 & \cellcolor[HTML]{21918c} \color[HTML]{f1f1f1} \ding{51} & \cellcolor[HTML]{440154} \color[HTML]{f1f1f1} \ding{55} & \cellcolor[HTML]{21918c} \color[HTML]{f1f1f1} \ding{51} & \cellcolor[HTML]{440154} \color[HTML]{f1f1f1} \ding{55} & \cellcolor[HTML]{440154} \color[HTML]{f1f1f1} \ding{55} & \cellcolor[HTML]{440154} \color[HTML]{f1f1f1} \ding{55} & \cellcolor[HTML]{440154} \color[HTML]{f1f1f1} \ding{55} & \cellcolor[HTML]{21918c} \color[HTML]{f1f1f1} \ding{51} & \cellcolor[HTML]{21918c} \color[HTML]{f1f1f1} \ding{51} & \cellcolor[HTML]{21918c} \color[HTML]{f1f1f1} \ding{51} & \cellcolor[HTML]{21918c} \color[HTML]{f1f1f1} \ding{51} & \cellcolor[HTML]{440154} \color[HTML]{f1f1f1} \ding{55} & \cellcolor[HTML]{21918c} \color[HTML]{f1f1f1} \ding{51} & \cellcolor[HTML]{21918c} \color[HTML]{f1f1f1} \ding{51} \\
{[AWS]}/20aa329b.../... & 5327 & \cellcolor[HTML]{21918c} \color[HTML]{f1f1f1} \ding{51} & \cellcolor[HTML]{440154} \color[HTML]{f1f1f1} \ding{55} & \cellcolor[HTML]{21918c} \color[HTML]{f1f1f1} \ding{51} & \cellcolor[HTML]{440154} \color[HTML]{f1f1f1} \ding{55} & \cellcolor[HTML]{440154} \color[HTML]{f1f1f1} \ding{55} & \cellcolor[HTML]{440154} \color[HTML]{f1f1f1} \ding{55} & \cellcolor[HTML]{440154} \color[HTML]{f1f1f1} \ding{55} & \cellcolor[HTML]{21918c} \color[HTML]{f1f1f1} \ding{51} & \cellcolor[HTML]{21918c} \color[HTML]{f1f1f1} \ding{51} & \cellcolor[HTML]{21918c} \color[HTML]{f1f1f1} \ding{51} & \cellcolor[HTML]{21918c} \color[HTML]{f1f1f1} \ding{51} & \cellcolor[HTML]{440154} \color[HTML]{f1f1f1} \ding{55} & \cellcolor[HTML]{21918c} \color[HTML]{f1f1f1} \ding{51} & \cellcolor[HTML]{21918c} \color[HTML]{f1f1f1} \ding{51} \\
{[AWS]}/f703696e.../... & 4592 & \cellcolor[HTML]{21918c} \color[HTML]{f1f1f1} \ding{51} & \cellcolor[HTML]{440154} \color[HTML]{f1f1f1} \ding{55} & \cellcolor[HTML]{21918c} \color[HTML]{f1f1f1} \ding{51} & \cellcolor[HTML]{440154} \color[HTML]{f1f1f1} \ding{55} & \cellcolor[HTML]{440154} \color[HTML]{f1f1f1} \ding{55} & \cellcolor[HTML]{440154} \color[HTML]{f1f1f1} \ding{55} & \cellcolor[HTML]{440154} \color[HTML]{f1f1f1} \ding{55} & \cellcolor[HTML]{21918c} \color[HTML]{f1f1f1} \ding{51} & \cellcolor[HTML]{21918c} \color[HTML]{f1f1f1} \ding{51} & \cellcolor[HTML]{21918c} \color[HTML]{f1f1f1} \ding{51} & \cellcolor[HTML]{21918c} \color[HTML]{f1f1f1} \ding{51} & \cellcolor[HTML]{440154} \color[HTML]{f1f1f1} \ding{55} & \cellcolor[HTML]{21918c} \color[HTML]{f1f1f1} \ding{51} & \cellcolor[HTML]{21918c} \color[HTML]{f1f1f1} \ding{51} \\
{[AWS]}/08c2f264.../... & 1136 & \cellcolor[HTML]{21918c} \color[HTML]{f1f1f1} \ding{51} & \cellcolor[HTML]{440154} \color[HTML]{f1f1f1} \ding{55} & \cellcolor[HTML]{21918c} \color[HTML]{f1f1f1} \ding{51} & \cellcolor[HTML]{440154} \color[HTML]{f1f1f1} \ding{55} & \cellcolor[HTML]{440154} \color[HTML]{f1f1f1} \ding{55} & \cellcolor[HTML]{440154} \color[HTML]{f1f1f1} \ding{55} & \cellcolor[HTML]{440154} \color[HTML]{f1f1f1} \ding{55} & \cellcolor[HTML]{21918c} \color[HTML]{f1f1f1} \ding{51} & \cellcolor[HTML]{21918c} \color[HTML]{f1f1f1} \ding{51} & \cellcolor[HTML]{21918c} \color[HTML]{f1f1f1} \ding{51} & \cellcolor[HTML]{21918c} \color[HTML]{f1f1f1} \ding{51} & \cellcolor[HTML]{440154} \color[HTML]{f1f1f1} \ding{55} & \cellcolor[HTML]{21918c} \color[HTML]{f1f1f1} \ding{51} & \cellcolor[HTML]{21918c} \color[HTML]{f1f1f1} \ding{51} \\
{[AWS]}/dba8f01c.../... & 868 & \cellcolor[HTML]{21918c} \color[HTML]{f1f1f1} \ding{51} & \cellcolor[HTML]{440154} \color[HTML]{f1f1f1} \ding{55} & \cellcolor[HTML]{21918c} \color[HTML]{f1f1f1} \ding{51} & \cellcolor[HTML]{440154} \color[HTML]{f1f1f1} \ding{55} & \cellcolor[HTML]{440154} \color[HTML]{f1f1f1} \ding{55} & \cellcolor[HTML]{440154} \color[HTML]{f1f1f1} \ding{55} & \cellcolor[HTML]{440154} \color[HTML]{f1f1f1} \ding{55} & \cellcolor[HTML]{21918c} \color[HTML]{f1f1f1} \ding{51} & \cellcolor[HTML]{21918c} \color[HTML]{f1f1f1} \ding{51} & \cellcolor[HTML]{21918c} \color[HTML]{f1f1f1} \ding{51} & \cellcolor[HTML]{21918c} \color[HTML]{f1f1f1} \ding{51} & \cellcolor[HTML]{440154} \color[HTML]{f1f1f1} \ding{55} & \cellcolor[HTML]{21918c} \color[HTML]{f1f1f1} \ding{51} & \cellcolor[HTML]{21918c} \color[HTML]{f1f1f1} \ding{51} \\
cloudie.rpki.app/rrdp/... & 396 & \cellcolor[HTML]{21918c} \color[HTML]{f1f1f1} \ding{51} & \cellcolor[HTML]{440154} \color[HTML]{f1f1f1} \ding{55} & \cellcolor[HTML]{21918c} \color[HTML]{f1f1f1} \ding{51} & \cellcolor[HTML]{440154} \color[HTML]{f1f1f1} \ding{55} & \cellcolor[HTML]{440154} \color[HTML]{f1f1f1} \ding{55} & \cellcolor[HTML]{21918c} \color[HTML]{f1f1f1} \ding{51} & \cellcolor[HTML]{440154} \color[HTML]{f1f1f1} \ding{55} & \cellcolor[HTML]{21918c} \color[HTML]{f1f1f1} \ding{51} & \cellcolor[HTML]{21918c} \color[HTML]{f1f1f1} \ding{51} & \cellcolor[HTML]{440154} \color[HTML]{f1f1f1} \ding{55} & \cellcolor[HTML]{440154} \color[HTML]{f1f1f1} \ding{55} & \cellcolor[HTML]{21918c} \color[HTML]{f1f1f1} \ding{51} & \cellcolor[HTML]{21918c} \color[HTML]{f1f1f1} \ding{51} & \cellcolor[HTML]{21918c} \color[HTML]{f1f1f1} \ding{51} \\
{[AWS]}/54602fb0.../... & 339 & \cellcolor[HTML]{21918c} \color[HTML]{f1f1f1} \ding{51} & \cellcolor[HTML]{440154} \color[HTML]{f1f1f1} \ding{55} & \cellcolor[HTML]{21918c} \color[HTML]{f1f1f1} \ding{51} & \cellcolor[HTML]{440154} \color[HTML]{f1f1f1} \ding{55} & \cellcolor[HTML]{440154} \color[HTML]{f1f1f1} \ding{55} & \cellcolor[HTML]{440154} \color[HTML]{f1f1f1} \ding{55} & \cellcolor[HTML]{440154} \color[HTML]{f1f1f1} \ding{55} & \cellcolor[HTML]{21918c} \color[HTML]{f1f1f1} \ding{51} & \cellcolor[HTML]{21918c} \color[HTML]{f1f1f1} \ding{51} & \cellcolor[HTML]{21918c} \color[HTML]{f1f1f1} \ding{51} & \cellcolor[HTML]{21918c} \color[HTML]{f1f1f1} \ding{51} & \cellcolor[HTML]{440154} \color[HTML]{f1f1f1} \ding{55} & \cellcolor[HTML]{21918c} \color[HTML]{f1f1f1} \ding{51} & \cellcolor[HTML]{21918c} \color[HTML]{f1f1f1} \ding{51} \\
{[AWS]}/517f3ed7.../... & 310 & \cellcolor[HTML]{21918c} \color[HTML]{f1f1f1} \ding{51} & \cellcolor[HTML]{440154} \color[HTML]{f1f1f1} \ding{55} & \cellcolor[HTML]{21918c} \color[HTML]{f1f1f1} \ding{51} & \cellcolor[HTML]{440154} \color[HTML]{f1f1f1} \ding{55} & \cellcolor[HTML]{440154} \color[HTML]{f1f1f1} \ding{55} & \cellcolor[HTML]{440154} \color[HTML]{f1f1f1} \ding{55} & \cellcolor[HTML]{440154} \color[HTML]{f1f1f1} \ding{55} & \cellcolor[HTML]{21918c} \color[HTML]{f1f1f1} \ding{51} & \cellcolor[HTML]{21918c} \color[HTML]{f1f1f1} \ding{51} & \cellcolor[HTML]{21918c} \color[HTML]{f1f1f1} \ding{51} & \cellcolor[HTML]{21918c} \color[HTML]{f1f1f1} \ding{51} & \cellcolor[HTML]{440154} \color[HTML]{f1f1f1} \ding{55} & \cellcolor[HTML]{21918c} \color[HTML]{f1f1f1} \ding{51} & \cellcolor[HTML]{21918c} \color[HTML]{f1f1f1} \ding{51} \\
rpki.admin.freerangecloud.com/rrdp/... & 221 & \cellcolor[HTML]{440154} \color[HTML]{f1f1f1} \ding{55} & \cellcolor[HTML]{440154} \color[HTML]{f1f1f1} \ding{55} & \cellcolor[HTML]{440154} \color[HTML]{f1f1f1} \ding{55} & \cellcolor[HTML]{440154} \color[HTML]{f1f1f1} \ding{55} & \cellcolor[HTML]{440154} \color[HTML]{f1f1f1} \ding{55} & \cellcolor[HTML]{21918c} \color[HTML]{f1f1f1} \ding{51} & \cellcolor[HTML]{440154} \color[HTML]{f1f1f1} \ding{55} & \cellcolor[HTML]{21918c} \color[HTML]{f1f1f1} \ding{51} & \cellcolor[HTML]{21918c} \color[HTML]{f1f1f1} \ding{51} & \cellcolor[HTML]{21918c} \color[HTML]{f1f1f1} \ding{51} & \cellcolor[HTML]{440154} \color[HTML]{f1f1f1} \ding{55} & \cellcolor[HTML]{21918c} \color[HTML]{f1f1f1} \ding{51} & \cellcolor[HTML]{21918c} \color[HTML]{f1f1f1} \ding{51} & \cellcolor[HTML]{21918c} \color[HTML]{f1f1f1} \ding{51} \\
magellan.ipxo.com/rrdp/... & 210 & \cellcolor[HTML]{21918c} \color[HTML]{f1f1f1} \ding{51} & \cellcolor[HTML]{440154} \color[HTML]{f1f1f1} \ding{55} & \cellcolor[HTML]{21918c} \color[HTML]{f1f1f1} \ding{51} & \cellcolor[HTML]{440154} \color[HTML]{f1f1f1} \ding{55} & \cellcolor[HTML]{440154} \color[HTML]{f1f1f1} \ding{55} & \cellcolor[HTML]{21918c} \color[HTML]{f1f1f1} \ding{51} & \cellcolor[HTML]{440154} \color[HTML]{f1f1f1} \ding{55} & \cellcolor[HTML]{21918c} \color[HTML]{f1f1f1} \ding{51} & \cellcolor[HTML]{21918c} \color[HTML]{f1f1f1} \ding{51} & \cellcolor[HTML]{21918c} \color[HTML]{f1f1f1} \ding{51} & \cellcolor[HTML]{21918c} \color[HTML]{f1f1f1} \ding{51} & \cellcolor[HTML]{21918c} \color[HTML]{f1f1f1} \ding{51} & \cellcolor[HTML]{21918c} \color[HTML]{f1f1f1} \ding{51} & \cellcolor[HTML]{21918c} \color[HTML]{f1f1f1} \ding{51} \\
{[AWS]}/967a255c.../... & 183 & \cellcolor[HTML]{21918c} \color[HTML]{f1f1f1} \ding{51} & \cellcolor[HTML]{440154} \color[HTML]{f1f1f1} \ding{55} & \cellcolor[HTML]{21918c} \color[HTML]{f1f1f1} \ding{51} & \cellcolor[HTML]{440154} \color[HTML]{f1f1f1} \ding{55} & \cellcolor[HTML]{440154} \color[HTML]{f1f1f1} \ding{55} & \cellcolor[HTML]{440154} \color[HTML]{f1f1f1} \ding{55} & \cellcolor[HTML]{440154} \color[HTML]{f1f1f1} \ding{55} & \cellcolor[HTML]{21918c} \color[HTML]{f1f1f1} \ding{51} & \cellcolor[HTML]{21918c} \color[HTML]{f1f1f1} \ding{51} & \cellcolor[HTML]{21918c} \color[HTML]{f1f1f1} \ding{51} & \cellcolor[HTML]{21918c} \color[HTML]{f1f1f1} \ding{51} & \cellcolor[HTML]{440154} \color[HTML]{f1f1f1} \ding{55} & \cellcolor[HTML]{21918c} \color[HTML]{f1f1f1} \ding{51} & \cellcolor[HTML]{21918c} \color[HTML]{f1f1f1} \ding{51} \\
{[AWS]}/e72d8db0.../... & 163 & \cellcolor[HTML]{21918c} \color[HTML]{f1f1f1} \ding{51} & \cellcolor[HTML]{440154} \color[HTML]{f1f1f1} \ding{55} & \cellcolor[HTML]{21918c} \color[HTML]{f1f1f1} \ding{51} & \cellcolor[HTML]{440154} \color[HTML]{f1f1f1} \ding{55} & \cellcolor[HTML]{440154} \color[HTML]{f1f1f1} \ding{55} & \cellcolor[HTML]{440154} \color[HTML]{f1f1f1} \ding{55} & \cellcolor[HTML]{440154} \color[HTML]{f1f1f1} \ding{55} & \cellcolor[HTML]{21918c} \color[HTML]{f1f1f1} \ding{51} & \cellcolor[HTML]{21918c} \color[HTML]{f1f1f1} \ding{51} & \cellcolor[HTML]{21918c} \color[HTML]{f1f1f1} \ding{51} & \cellcolor[HTML]{21918c} \color[HTML]{f1f1f1} \ding{51} & \cellcolor[HTML]{440154} \color[HTML]{f1f1f1} \ding{55} & \cellcolor[HTML]{21918c} \color[HTML]{f1f1f1} \ding{51} & \cellcolor[HTML]{21918c} \color[HTML]{f1f1f1} \ding{51} \\
rpki.roa.net/rrdp/... & 150 & \cellcolor[HTML]{440154} \color[HTML]{f1f1f1} \ding{55} & \cellcolor[HTML]{440154} \color[HTML]{f1f1f1} \ding{55} & \cellcolor[HTML]{440154} \color[HTML]{f1f1f1} \ding{55} & \cellcolor[HTML]{21918c} \color[HTML]{f1f1f1} \ding{51} & \cellcolor[HTML]{21918c} \color[HTML]{f1f1f1} \ding{51} & \cellcolor[HTML]{21918c} \color[HTML]{f1f1f1} \ding{51} & \cellcolor[HTML]{440154} \color[HTML]{f1f1f1} \ding{55} & \cellcolor[HTML]{21918c} \color[HTML]{f1f1f1} \ding{51} & \cellcolor[HTML]{21918c} \color[HTML]{f1f1f1} \ding{51} & \cellcolor[HTML]{21918c} \color[HTML]{f1f1f1} \ding{51} & \cellcolor[HTML]{440154} \color[HTML]{f1f1f1} \ding{55} & \cellcolor[HTML]{21918c} \color[HTML]{f1f1f1} \ding{51} & \cellcolor[HTML]{21918c} \color[HTML]{f1f1f1} \ding{51} & \cellcolor[HTML]{21918c} \color[HTML]{f1f1f1} \ding{51} \\
krill.47272.net/rrdp/... & 129 & \cellcolor[HTML]{440154} \color[HTML]{f1f1f1} \ding{55} & \cellcolor[HTML]{440154} \color[HTML]{f1f1f1} \ding{55} & \cellcolor[HTML]{440154} \color[HTML]{f1f1f1} \ding{55} & \cellcolor[HTML]{440154} \color[HTML]{f1f1f1} \ding{55} & \cellcolor[HTML]{440154} \color[HTML]{f1f1f1} \ding{55} & \cellcolor[HTML]{21918c} \color[HTML]{f1f1f1} \ding{51} & \cellcolor[HTML]{440154} \color[HTML]{f1f1f1} \ding{55} & \cellcolor[HTML]{21918c} \color[HTML]{f1f1f1} \ding{51} & \cellcolor[HTML]{440154} \color[HTML]{f1f1f1} \ding{55} & \cellcolor[HTML]{440154} \color[HTML]{f1f1f1} \ding{55} & \cellcolor[HTML]{21918c} \color[HTML]{f1f1f1} \ding{51} & \cellcolor[HTML]{21918c} \color[HTML]{f1f1f1} \ding{51} & \cellcolor[HTML]{21918c} \color[HTML]{f1f1f1} \ding{51} & \cellcolor[HTML]{21918c} \color[HTML]{f1f1f1} \ding{51} \\
rpki-01.pdxnet.uk/rrdp/... & 119 & \cellcolor[HTML]{440154} \color[HTML]{f1f1f1} \ding{55} & \cellcolor[HTML]{440154} \color[HTML]{f1f1f1} \ding{55} & \cellcolor[HTML]{440154} \color[HTML]{f1f1f1} \ding{55} & \cellcolor[HTML]{21918c} \color[HTML]{f1f1f1} \ding{51} & \cellcolor[HTML]{21918c} \color[HTML]{f1f1f1} \ding{51} & \cellcolor[HTML]{21918c} \color[HTML]{f1f1f1} \ding{51} & \cellcolor[HTML]{440154} \color[HTML]{f1f1f1} \ding{55} & \cellcolor[HTML]{21918c} \color[HTML]{f1f1f1} \ding{51} & \cellcolor[HTML]{440154} \color[HTML]{f1f1f1} \ding{55} & \cellcolor[HTML]{440154} \color[HTML]{f1f1f1} \ding{55} & \cellcolor[HTML]{440154} \color[HTML]{f1f1f1} \ding{55} & \cellcolor[HTML]{21918c} \color[HTML]{f1f1f1} \ding{51} & \cellcolor[HTML]{21918c} \color[HTML]{f1f1f1} \ding{51} & \cellcolor[HTML]{21918c} \color[HTML]{f1f1f1} \ding{51} \\
{[AWS]}/fe3737fb.../... & 108 & \cellcolor[HTML]{21918c} \color[HTML]{f1f1f1} \ding{51} & \cellcolor[HTML]{440154} \color[HTML]{f1f1f1} \ding{55} & \cellcolor[HTML]{21918c} \color[HTML]{f1f1f1} \ding{51} & \cellcolor[HTML]{440154} \color[HTML]{f1f1f1} \ding{55} & \cellcolor[HTML]{440154} \color[HTML]{f1f1f1} \ding{55} & \cellcolor[HTML]{440154} \color[HTML]{f1f1f1} \ding{55} & \cellcolor[HTML]{440154} \color[HTML]{f1f1f1} \ding{55} & \cellcolor[HTML]{21918c} \color[HTML]{f1f1f1} \ding{51} & \cellcolor[HTML]{21918c} \color[HTML]{f1f1f1} \ding{51} & \cellcolor[HTML]{21918c} \color[HTML]{f1f1f1} \ding{51} & \cellcolor[HTML]{21918c} \color[HTML]{f1f1f1} \ding{51} & \cellcolor[HTML]{440154} \color[HTML]{f1f1f1} \ding{55} & \cellcolor[HTML]{21918c} \color[HTML]{f1f1f1} \ding{51} & \cellcolor[HTML]{21918c} \color[HTML]{f1f1f1} \ding{51} \\
ca.nat.moe/rrdp/... & 99 & \cellcolor[HTML]{440154} \color[HTML]{f1f1f1} \ding{55} & \cellcolor[HTML]{440154} \color[HTML]{f1f1f1} \ding{55} & \cellcolor[HTML]{440154} \color[HTML]{f1f1f1} \ding{55} & \cellcolor[HTML]{21918c} \color[HTML]{f1f1f1} \ding{51} & \cellcolor[HTML]{21918c} \color[HTML]{f1f1f1} \ding{51} & \cellcolor[HTML]{21918c} \color[HTML]{f1f1f1} \ding{51} & \cellcolor[HTML]{440154} \color[HTML]{f1f1f1} \ding{55} & \cellcolor[HTML]{21918c} \color[HTML]{f1f1f1} \ding{51} & \cellcolor[HTML]{21918c} \color[HTML]{f1f1f1} \ding{51} & \cellcolor[HTML]{440154} \color[HTML]{f1f1f1} \ding{55} & \cellcolor[HTML]{440154} \color[HTML]{f1f1f1} \ding{55} & \cellcolor[HTML]{21918c} \color[HTML]{f1f1f1} \ding{51} & \cellcolor[HTML]{21918c} \color[HTML]{f1f1f1} \ding{51} & \cellcolor[HTML]{21918c} \color[HTML]{f1f1f1} \ding{51} \\
{[AWS]}/b8a1dd25.../... & 89 & \cellcolor[HTML]{21918c} \color[HTML]{f1f1f1} \ding{51} & \cellcolor[HTML]{440154} \color[HTML]{f1f1f1} \ding{55} & \cellcolor[HTML]{21918c} \color[HTML]{f1f1f1} \ding{51} & \cellcolor[HTML]{440154} \color[HTML]{f1f1f1} \ding{55} & \cellcolor[HTML]{440154} \color[HTML]{f1f1f1} \ding{55} & \cellcolor[HTML]{440154} \color[HTML]{f1f1f1} \ding{55} & \cellcolor[HTML]{440154} \color[HTML]{f1f1f1} \ding{55} & \cellcolor[HTML]{21918c} \color[HTML]{f1f1f1} \ding{51} & \cellcolor[HTML]{21918c} \color[HTML]{f1f1f1} \ding{51} & \cellcolor[HTML]{21918c} \color[HTML]{f1f1f1} \ding{51} & \cellcolor[HTML]{21918c} \color[HTML]{f1f1f1} \ding{51} & \cellcolor[HTML]{440154} \color[HTML]{f1f1f1} \ding{55} & \cellcolor[HTML]{21918c} \color[HTML]{f1f1f1} \ding{51} & \cellcolor[HTML]{21918c} \color[HTML]{f1f1f1} \ding{51} \\
rrdp.rp.ki/... & 76 & \cellcolor[HTML]{21918c} \color[HTML]{f1f1f1} \ding{51} & \cellcolor[HTML]{440154} \color[HTML]{f1f1f1} \ding{55} & \cellcolor[HTML]{21918c} \color[HTML]{f1f1f1} \ding{51} & \cellcolor[HTML]{21918c} \color[HTML]{f1f1f1} \ding{51} & \cellcolor[HTML]{21918c} \color[HTML]{f1f1f1} \ding{51} & \cellcolor[HTML]{21918c} \color[HTML]{f1f1f1} \ding{51} & \cellcolor[HTML]{440154} \color[HTML]{f1f1f1} \ding{55} & \cellcolor[HTML]{21918c} \color[HTML]{f1f1f1} \ding{51} & \cellcolor[HTML]{21918c} \color[HTML]{f1f1f1} \ding{51} & \cellcolor[HTML]{440154} \color[HTML]{f1f1f1} \ding{55} & \cellcolor[HTML]{21918c} \color[HTML]{f1f1f1} \ding{51} & \cellcolor[HTML]{21918c} \color[HTML]{f1f1f1} \ding{51} & \cellcolor[HTML]{21918c} \color[HTML]{f1f1f1} \ding{51} & \cellcolor[HTML]{21918c} \color[HTML]{f1f1f1} \ding{51} \\
repo.rpki.space/rrdp/... & 75 & \cellcolor[HTML]{440154} \color[HTML]{f1f1f1} \ding{55} & \cellcolor[HTML]{440154} \color[HTML]{f1f1f1} \ding{55} & \cellcolor[HTML]{440154} \color[HTML]{f1f1f1} \ding{55} & \cellcolor[HTML]{440154} \color[HTML]{f1f1f1} \ding{55} & \cellcolor[HTML]{440154} \color[HTML]{f1f1f1} \ding{55} & \cellcolor[HTML]{21918c} \color[HTML]{f1f1f1} \ding{51} & \cellcolor[HTML]{440154} \color[HTML]{f1f1f1} \ding{55} & \cellcolor[HTML]{21918c} \color[HTML]{f1f1f1} \ding{51} & \cellcolor[HTML]{21918c} \color[HTML]{f1f1f1} \ding{51} & \cellcolor[HTML]{21918c} \color[HTML]{f1f1f1} \ding{51} & \cellcolor[HTML]{21918c} \color[HTML]{f1f1f1} \ding{51} & \cellcolor[HTML]{21918c} \color[HTML]{f1f1f1} \ding{51} & \cellcolor[HTML]{21918c} \color[HTML]{f1f1f1} \ding{51} & \cellcolor[HTML]{21918c} \color[HTML]{f1f1f1} \ding{51} \\
\end{tabular}
\caption{Implemented BCPs by 20 largest repositories by ROA count operated by other instances}
\label{table:bcps_20l}
\end{table}

\begin{figure}
  \centering
  \includegraphics[width=0.5\textwidth]{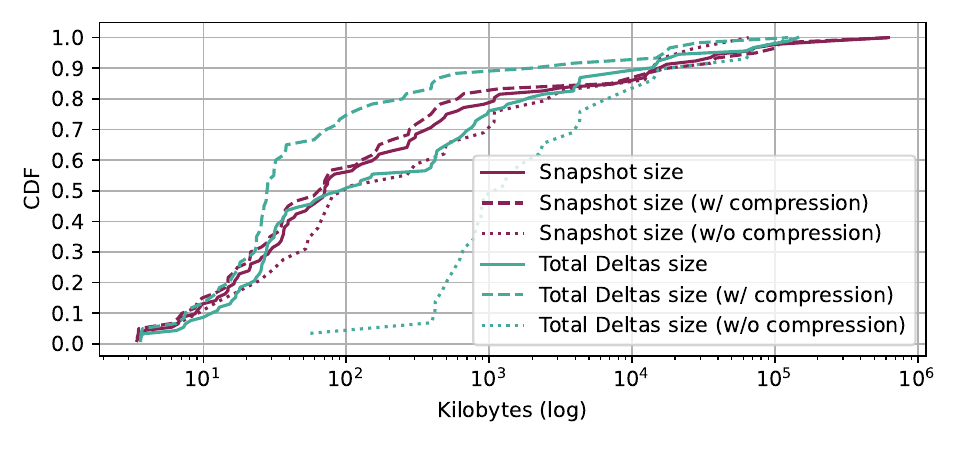}
  \caption{Maximum snapshot and delta file size per repository over the course of 6 days served by all RRDP servers, and servers that support and do not support compression.}
  \label{fig:snapshot_delta_cdf}
\end{figure}

\begin{acronym}
\acro{AS}{Autonomous System}
\acrodefplural{AS}[ASes]{Autonomous Systems}
\acro{ASN}{Autonomous System Number}
\acro{ASPA}{Autonomous System Provider Authorisation}
\acro{AWS}{Amazon Web Services}

\acro{BGP}{Border Gateway Protocol}
\acro{BCP}{Best Current Practice}
\acrodefplural{BCP}{Best Current Practices}

\acro{CA}{Certificate Authority}
\acrodefplural{CA}{Certificate Authorities}
\acro{CDN}{Content Delivery Network}
\acro{CSP}{Cloud Service Provider}
\acrodefplural{CSP}{Cloud Service Providers}
\acro{CRL}{Certificate Revocation List}

\acro{DNS}{Domain Name System}
\acro{DNSSEC}{Domain Name System Security}

\acro{FQDN}{Fully Qualified Domain Name}

\acro{IETF}{Internet Engineering Task Force}
\acro{IXP}{Internet Exchange Point}

\acro{NIR}{National Internet Registry}
\acrodefplural{NIR}{National Internet Registries}
\acro{NLRI}{Network Layer Reachability Information}

\acro{RIR}{Regional Internet Registry}
\acrodefplural{RIR}{Regional Internet Registries}
\acro{RP}{Relying Party}
\acrodefplural{RP}{Relying Parties}
\acro{ROA}{Route Origin Authorisation}
\acrodefplural{ROA}{Route Origin Authorisations}
\acro{ROV}{Route Origin Validation}
\acro{RPKI}{Resource Public Key Infrastructure}
\acro{RSC}{RPKI Signed Checklist}
\acrodefplural{RSC}{RPKI Signed Checklists}
\acro{RRDP}{RPKI Repository Delta Protocol}

\acro{TLD}{Top-Level Domain}

\acro{URI}{Uniform Resource Identifier}
\acro{URL}{Uniform Resource Locator}
\acrodefplural{URL}{Uniform Resource Locators}
\end{acronym}

\end{appendices}

\end{document}